\documentclass[10pt,twocolumn]{IEEEtran}
\usepackage{multicol}

\usepackage{color}
\usepackage{soul}
\usepackage{amsmath}
\usepackage{graphics}
\usepackage{setspace}
\usepackage{cite}
\usepackage{latexsym}
\usepackage{float}
\usepackage{epsfig}
\usepackage{multirow}
\usepackage[table,xcdraw]{xcolor}
\usepackage{cite,cases,url}
\usepackage{amssymb}
\usepackage{graphicx}
\usepackage{fancyhdr}
\usepackage{epstopdf}
\usepackage{balance}
\usepackage{subcaption}
\usepackage{enumerate}
\usepackage{array}
\usepackage{algorithmic}
\usepackage{algorithm}
\usepackage{enumitem}
\usepackage[normalem]{ulem}
\useunder{\uline}{\ul}{}

\newcolumntype{L}{>{\centering\arraybackslash}m{5cm}}
\newcolumntype{K}{>{\centering\arraybackslash}m{6cm}}
\newcolumntype{P}{>{\centering\arraybackslash}m{2.3cm}}
\newcolumntype{M}{>{\raggedright\arraybackslash}m{2cm}}
\newcolumntype{N}{>{\raggedright\arraybackslash}m{2.5cm}}

\fancypagestyle{firstpage}
{
    \fancyhead[L]{\footnotesize This article has been accepted for publication in the IEEE Wireless Communications Magazine.  Please cite it as A. S. Abdalla, J. Moore, N. Adihikari, and V. Marojevic, ''ZTRAN: Prototyping Zero Trust Security xApps for Open Radio Access Network Deployments,'' IEEE Wireless Communications Magazine, April 2024. } 
    \fancyhead[R]{\footnotesize \thepage}
    \cfoot{}
    
}

\begin{document}

\title{
ZTRAN: Prototyping Zero Trust Security xApps for Open Radio Access Network Deployments}
\author{
\IEEEauthorblockN{Aly S. Abdalla, Joshua Moore, Nisha Adhikari, and Vuk Marojevic \\
} 
\normalsize\IEEEauthorblockA{Dept. of Electrical and Computer Engineering,  Mississippi State University, USA}\\
\normalsize\IEEEauthorblockA{Emails:  \{asa298; jjm702; na731;   vuk.marojevic\}@msstate.edu}
}
\maketitle
\thispagestyle{firstpage}
\begin{abstract}
\textcolor{black}{The open radio access network (O-RAN) offers new degrees of 
freedom for building and operating advanced cellular networks. Emphasizing on RAN disaggregation, open interfaces, multi-vendor support, and RAN intelligent controllers (RICs), O-RAN facilitates adaptation to new applications and technology trends. Yet, this architecture introduces new security challenges. 
This paper 
proposes leveraging zero trust principles 
for O-RAN security. 
We introduce zero trust RAN (ZTRAN), which embeds 
service authentication, intrusion detection, and secure slicing subsystems that are encapsulated as xApps. 
We implement ZTRAN on the open artificial intelligence cellular (OAIC) research platform and demonstrate its feasibility and effectiveness in terms of legitimate user throughput and latency figures. Our experimental analysis illustrates how ZTRAN's intrusion detection and secure slicing microservices operate effectively and in concert as part of O-RAN Alliance's containerized near-real time RIC. 
Research directions include exploring machine learning and additional threat intelligence feeds for improving the performance and extending the scope of ZTRAN. 
}
\end{abstract}
\IEEEpeerreviewmaketitle

\section{Introduction}
\label{sec:intro}
\textcolor{black}{
The open radio access network (O-RAN) 
is revolutionizing the cellular industry by enhancing flexibility, interoperability, and cost efficiency. Unlike traditional networks that rely on proprietary and tightly integrated solutions, 
}
\textcolor{black}{
O-RAN promotes 
vendor neutrality. 
It eliminates 
the confinement of network operators to specific vendors by advocating for open interfaces and standardized implementations to facilitate interoperability between 
the hardware and software from different vendors~\cite{Aly_ORAN}. This encourages healthy competition and drives industry innovations. 
The disaggregated and software-defined nature of O-RAN allows network operators to leverage 
commercial off-the-shelf hardware and virtualization technology, leading to reduced capital and operational expenditures. 
} 

\textcolor{black}{O-RAN's modular architecture, introducing RAN intelligent controllers (RICs) for flexible network management, empowers network operators to swiftly adapt to new application or service demands and to emerging technology trends. This facilitates easy integration of new features, functionalities, and network enhancements, ensuring networks remain resilient and capable of supporting future needs.}


\textcolor{black}{Network security is a critical aspect of emerging wireless networks, including O-RAN. 
The increasing sophistication of cyberthreats and the continuously evolving threat landscape demand proactive countermeasures. 
The conventional network security model follows a perimeter-based approach, assuming that once users are given access to the network, they can be trusted. 
However, this approach has proven inadequate against sophisticated cyberthreats and the 
expanding threat surface of open, interoperable, and virtualized wireless networks~\cite{abdalla2023end}. 
By adopting a zero trust approach to O-RAN security, network operators can enforce precise access control mechanisms based on user identities (IDs), continuously 
re-authenticate users, and closely monitor network activities for anomalous behavior. 
}

\textcolor{black}{
Some wireless network users demand high throughput, others low latency, and yet others secure and resilient communications, to name a few. This paper targets the latter user groups. Mission-critical users of cellular networks have specific quality of service (QoS) 
targets such as service availability, link reliability, and data privacy.} \textcolor{black}{The goal of this research is providing a comprehensive and detailed exploration of zero trust security methods, components, and their application 
to O-RAN. 
\textcolor{black}{We introduce the zero trust RAN (ZTRAN), a security framework composed of three microservices that embed security monitoring and control principles: service authentication, intrusion detection, and secure slicing. These microservices are hosted in O-RAN's near-real time (near-RT) RIC.}
Leveraging the Open Artificial Intelligence Cellular (OAIC) research platform\footnote{https://openaicellular.org}, we implement and 
experimentally demonstrate the practicality and effectiveness of 
ZTRAN and evaluate 
its challenges and limitations in addressing the security challenges specific to O-RAN deployments. }


\textcolor{black}{The rest of the paper is organized as follows: \textcolor{black}{Section II introduces 
zero trust principles and major network security components and discusses their potential application to 
O-RAN.} Section III  
introduces the ZTRAN components and processes. 
Section IV provides experimental results and analyses of our implementation of ZTRAN on the OAIC platform. Section V discusses critical open issues and research and development (R\&D) directions for
expanding ZTRAN's capabilities and security services within the near-RT RIC. Section VI offers the concluding remarks.}

\section{
\textcolor{black}{
Zero Trust Security and the O-RAN Use Case}} 
\label{sec:services}
\textcolor{black}{
Zero trust networks operate 
on the principle that the system never trusts and always verifies 
any request or action. It challenges the inherent assumption of trust within the network and adopts a more proactive and cautious stance. Zero trust 
access control verifies 
every access request, irrespective of the user or request. 
Under the zero trust 
security principles, every user, device, and application attempting to access the network is treated as potentially untrustworthy, regardless of their location or previous access privileges. 
This is on the basis that threats can emerge from both external and internal sources, and even trusted entities can become compromised. Zero trust emphasizes the need for continuous verification and authentication throughout the network, constantly scrutinizing and validating the IDs, security 
configurations, and behaviors of network users and devices~\cite{Ztrust1}. By adopting the zero trust model, communication networks can be protected by applying strong authentication methods, leveraging network segmentation, and simplifying granular access policies.}

\textcolor{black}{One part of a zero trust security chain is often 
a comprehensive monitoring and logging system that enables the swift detection of anomalous behaviors and 
security incidents. By closely monitoring network traffic, analyzing user activity, and correlating data from various sources, 
operators can 
identify potential threats and respond effectively to mitigate larger risks. Zero trust emphasizes the importance of having visibility and control over endpoints. This involves deploying endpoint security solutions that provide real time visibility into device health, compliance status, and security posture. Endpoint control mechanisms, such as enforcing security configurations, patch management, and software whitelisting, help ensure that endpoints connecting to the network are secure and meet the organization's security standards~\cite{Ztrust}.}

\begin{figure}[t]
    \centering
    \includegraphics[width=0.48\textwidth]{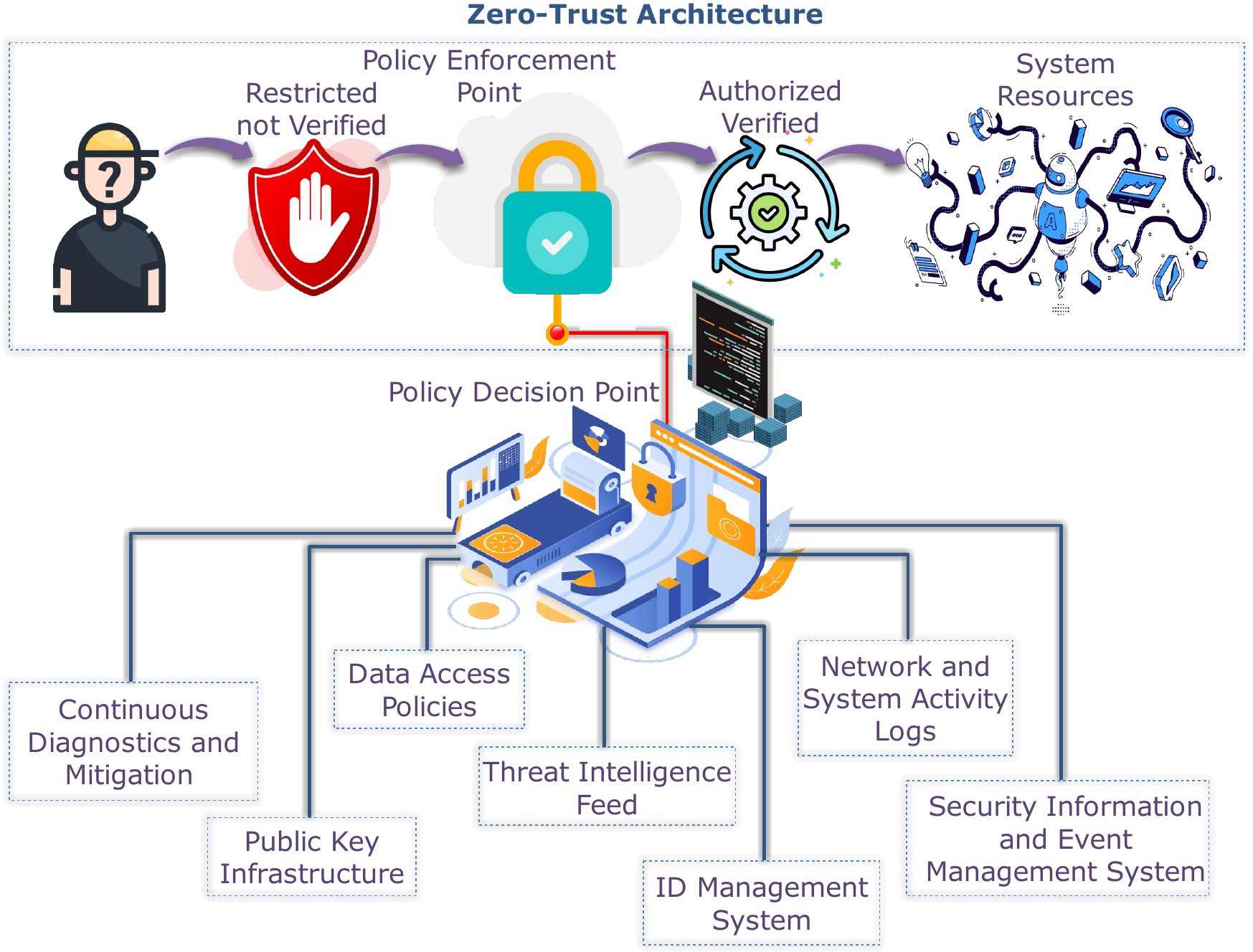}
    \vspace{-6 pt}
    \caption{Zero-trust security components. 
    } 
    \label{fig:ZeroTrustARCH}
    \vspace{-7mm}
\end{figure}

\textcolor{black}{
Two logical components enable the above zero thrust functionalities. 
These are 
the policy enforcement point (PEP) and the policy decision point (PDP) which can be 
operated on-premise 
or through a Cloud-based service~\cite{stafford2020zero}. Figure~\ref{fig:ZeroTrustARCH} depicts the conceptual framework that illustrates the 
relationships and interactions between these components. 
}

\subsection{PEP and PDP}
\textcolor{black}{
The PEP serves as a vital logical component that is acting as a gatekeeper for communication paths between users, devices, or other entities and 
service providers for the purpose of managing, monitoring, and controlling ongoing and outgoing connections.  Its primary function is enforcing access policies and 
ensuring that only authorized entities can access 
resources. The PEP interacts with the 
PDP to forward access requests and receive policy updates. The PDP is responsible for making the final access decision for a given subject trying to access a resource. It takes actions based on that decision, which may 
grant, deny, or revoke access to the resource. If the access request is approved, the PDP establishes the session-specific authentication tokens or credentials for secure network access. 
On the other hand, if the request is denied or if previously approved sessions need to be revoked, the PDP instructs the PEP to terminate the established connection, effectively preventing the entity from accessing the resource. The PDP bases its decision on 
the system's access policies and other inputs from external sources, including different PDP subsystems.  }

\subsection{PDP Subsystems}
The PDP subsystems are composed of a continuous diagnostics and mitigation system, 
data access policies, threat intelligence feeds, the ID management system, the public key infrastructure (PKI), network and system activity logs, and the security information and event management (SIEM) system~\cite{stafford2020zero}, as shown in Fig.~\ref{fig:ZeroTrustARCH} and described in continuation. 
\begin{itemize}[left=2 pt]

 \item 
\textcolor{black}{\textbf{Continuous diagnostics and mitigation} plays a pivotal role by gathering up-to-date information about the current state of 
system assets. 
The primary responsibility of this system is to implement critical updates to configurations and software components.} 
\textcolor{black}{It assesses whether the assets are running adequately patched components, verifies the integrity of approved software components, identifies any non-approved components, and flags potential vulnerabilities.} 
\vspace{+1 mm}

\item 
\textcolor{black}{\textbf{Threat intelligence feeds} are a vital resource for a zero-trust security architecture. They provide valuable information from diverse sources, both internal and external, assisting the 
zero trust PDP in making informed access decisions. These feeds constantly update the PDP about emerging threats, newly discovered vulnerabilities, malware reports, and recent attacks on other assets. By leveraging this intelligence, the PDP can swiftly identify potential risks and prevent access from suspicious sources.} 
\vspace{+1 mm}
\item 
\textcolor{black}{\textbf{Network and system activity logs} serve as a valuable repository of security-centric data. These logs aggregate 
information, including network traffic, resource access actions, and system events. Providing real time or near-real time feedback, this data empowers 
the network to closely monitor 
and analyze potential threats.} 
\vspace{+1 mm}
\item 
\textcolor{black}{\textbf{Data access policies}, either statically encoded or dynamically generated by the PDP, outline the attributes, rules, and permissions for granting authorized access to components and applications within the 
system. The PKI is responsible for generating and managing certificates issued to various entities within the 
network. These certificates authenticate and secure resources, subjects, services, and applications. Additionally, the zero trust PKI may collaborate with external PKI 
systems to ensure a seamless security infrastructure.} 
\vspace{-3 mm}
\item 
\textcolor{black}{\textbf{The ID management system} is an essential component responsible for creating, storing, and managing user accounts and identity records. Using technologies such as the Lightweight Directory Access Protocol (LDAP), this system contains critical 
information along with role assignments and access attributes.} 
\vspace{+1 mm}
\item 
\textcolor{black}{\textbf{A state-of-the-art 
SIEM system} enables in-depth analysis of security events. This data-driven approach empowers 
systems to refine their policies and swiftly respond to potential attacks, ensuring robust defense against emerging threats.
}
\end{itemize}

\subsection{\textcolor{black}{O-RAN Security Needs and Opportunities}}
\textcolor{black}{The O-RAN architecture adds new interfaces and RIC services to traditional RANs. It is the openness and enhanced deployment and management flexibility of O-RAN, that presents various security challenges. Security needs to be managed across a 
disaggregated RAN, which may involve 
components from multiple vendors
~\cite{abdalla2023end}. 
The O-RAN building blocks---the O-RAN central, distributed, and radio units (O-CUs, O-DUs, O-RUs), the near and non-RT RIC, xApps and rApps, and the open interfaces---add new challenges to user/network/service authentication and access control, data confidentiality, and resource integrity~\cite{abdalla2023end}. Table~I presents a comprehensive overview of each component's vulnerabilities and the proposed mitigation strategies. The threat surface spans unauthorized access attempts through various O-RAN interfaces, such as the open fronthaul, 
insecure or conflicting xApps, faulty or non-complying O-CU, O-DU, O-RU, or RIC subsystems. 
}

\textcolor{black}{The mitigation strategies presented in Table~I incorporate essential principles of zero trust security, ensuring a proactive and multi-layered approach to effectively address the O-RAN security challenges. 
Zero trust practices can be leveraged to add an additional layer of security for preventing 
unauthorized access to O-RAN resources and mitigating the associated risks. 
In the dynamic and evolving landscape of O-RAN, continuous monitoring and adaptation are critical for ensuring service availability and maintaining diverse 
QoS levels of commercial and mission-critical users, among others.} 
\begin{table*}
\centering
\caption{O-RAN security threats and mitigation strategies. 
}
\footnotesize
\label{tab:TableORANSEC}
\centering
{\begin{tabular}{|p{2.4cm}|p{6.6cm}|p{7.8cm}|}
\hline
\textbf{O-RAN Component}  &\textbf{Security Challenges  } &   \textbf{\textcolor{black}{Proposed Mitigation Strategies
}} 
\\ \hline
Fronthaul 
& 
Unauthorized access to the fronthaul interface can result in data interception and manipulation, leading to service disruptions and unauthorized access to critical components in the network.
&  
\textcolor{black}{Use secure communication protocols to encrypt data, 
implement robust authentication mechanisms, 
regularly update cryptographic keys and certificates, and 
implement network monitoring and anomaly detection systems as a \textbf{continuous diagnostics and mitigation} practice of zero trust security.}\\
\hline
Near-RT RIC and xApps
& 
Malicious applications (Apps) 
targeting the near-RT RIC can access sensitive user information including user location and priority, manipulate network behavior, and compromise user privacy. Unauthorized access to RIC APIs can lead to unauthorized control over network resources. xApps deployed within the near-RT RICs can introduce vulnerabilities if not properly developed or configured. Exploiting xApp vulnerabilities can lead to data leaks and system instability. Unauthorized access to critical components can disrupt network operations. 
& 
\textcolor{black}{Implement secure authentication mechanisms for App access to the RIC APIs, 
employ \textbf{data access policies} such as fine-grained access controls with role-based access control (RBAC) 
to restrict App access rights, 
regularly audit and monitor App behavior through anomaly detection as the \textbf{continuous diagnostics and mitigation} solution 
for identifying and responding to suspicious activities, and 
apply encryption for protecting sensitive data transmitted between the RIC and Apps. \textcolor{black}{Implement input validation and parameterized queries that can be verified with the help of \textbf{threat intelligence feeds} 
and implement secure communication protocols to ensure xApp data confidentiality and integrity.}}\\
\hline
Non-RT RIC and rApps 
& 
Similar to the near-RT RIC and xApp threats, compromising non-RT RIC and rApp integrity can lead to service disruption and unauthorized access to non-real time optimization functions. 
& 
\textcolor{black}{Apply access controls and integrate them with \textbf{data access policies} 
to restrict communication and access to the non-RT RIC, employ secure communication protocols, implement \textbf{continuous diagnostics and mitigation} in the form of intrusion detection and prevention, 
and 
conduct regular security assessments and penetration testing.}\\
\hline
O-CU 
&  
Spoofing control plane (C-plane) messages can lead to unauthorized control, service disruption, and network misconfiguration. 
& 
\textcolor{black}{Implement message authentication and integrity checks for both downlink and uplink C-plane messages,  
employ secure communication protocols, and 
apply \textbf{data access policies} such as rate-limiting mechanisms on the C-plane interface.
} \\
\hline
O-DU 
& 
Unauthorized access to the O-DU C-plane can lead to C-plane manipulation, data breaches, and unauthorized access to the network. 
& 
\textcolor{black}{Implement strong access controls and secure authentication mechanisms, apply encryption for C-plane communication, employ message authentication mechanisms, monitor C-plane traffic for anomalies 
implementing \textbf{continuous diagnostics and mitigation}, 
and perform regular security audits and vulnerability assessments 
through a \textbf{state-of-the-art SIEM system}.}\\
\hline
O-RU 
& 
Compromising O-RU firmware or software can lead to unauthorized access, denial of service attacks, and degradation of radio signal processing. 
&  
\textcolor{black}{Implement strict access controls as \textbf{data access policies} 
for O-RU management interfaces and 
conduct regular security assessments.}\\
\hline
M-Plane 
& 
The M-Plane handles communication between O-DU and O-CU for management and control purposes. Man-in-the-middle attacks can lead to data interception, unauthorized access, and timing manipulation, impacting network operations. 
& 
\textcolor{black}{Implement mutual authentication and encryption for M-Plane communication, 
use secure communication protocols, 
regularly validate timing packets, 
implement secure key management, and 
monitor M-Plane traffic through a \textbf{continuous diagnostics and mitigation system} for real time anomaly detection.} \\
\hline
C-Plane 
& 
The C-Plane is responsible for controlling radio resources and network configurations. Spoofing of C-plane messages can lead to unauthorized control, service disruption, and network misconfiguration. & 
\textcolor{black}{Implement message authentication and integrity checks for C-plane communication by incorporating \textbf{threat intelligence feeds}, 
employ secure communication protocols, 
apply rate-limiting mechanisms as \textbf{data access policies} on the C-plane interface.} \\
\hline
U-Plane 
& 
The user plane (U-Plane) carries user data between the O-RU and the O-DU and also between the O-DU and O-CU. Intercepting U-Plane communications can lead to 
unauthorized access to user data or cause service disruption.
& 
\textcolor{black}{Implement secure communication channels, mutual authentication, and encryption for U-Plane communication, 
use secure protocols for data in transit, 
employ intrusion detection and prevention for \textbf{continuous diagnostics and mitigation} of U-Plane traffic manipulations, and  
implement secure key management for cryptographic keys as an 
\textbf{ID management system} for secure U-Plane communication.}\\
\hline
Machine Learning (ML) 
&  
Poisoning ML training data and altering ML models can lead to misleading results, privacy breaches, and unauthorized control over data driven decision-making. 
& 
\textcolor{black}{Perform data validation and anomaly detection during ML training by a collaborative \textbf{state-of-the-art SIEM system} and \textbf{threat intelligence feeds}, 
use model encryption and secure enclaves, 
regularly audit the performance of ML models to detect deviations from expected behavior, 
implement 
\textbf{data access policies} and authentication mechanisms for ML components, and 
monitor ML model inputs and outputs 
for suspicious activities, data, and potential privacy breaches.}\\
\hline
\end{tabular}
\vspace{-3 mm}
}
\end{table*}

\section{
Zero Trust RAN}
\label{sec:services}

The ZTRAN framework is composed of three microservices: service authentication, intrusion detection, and secure slicing xApps. 
  
\subsection{ZTRAN--Authentication xApp}

\textcolor{black}{The authentication xApp 
is responsible for ensuring the secure identification and verification of user equipment (UE) requesting access to O-RAN services. 
ZTRAN implements multi-factor authentication (MFA)
~\cite{Dasgupta2017}, which enhances the level of security 
as multiple forms of verification, identification, or credentials 
are 
processed~\cite{multi_factorAuth}.} 
\begin{table*}[ht]
\centering
\caption{
ZTRAN's security features, functionalities, inference host, and inference data. 
}
\footnotesize
{\begin{tabular}{|p{2.3cm}|p{6.8cm}|p{1.8cm}|p{3.0cm}|}
\hline
 \textbf{
 Security feature} 
 &  \textbf{
 Functionality description}  & \textbf{Inference host }  &   \textbf{Inference data}  

\\ \hline
\textcolor{black}{Authentication}
& \textcolor{black}{
Employs layered multi-factor authentication to 
establish the legitimacy of network users 
before providing 
O-RAN resources and services 
}
& Near-RT RIC   
& \textcolor{black}{Knowledge-based, possession-based, and biometric-based factors}
\\ \hline
\textcolor{black}{Intrusion detection
} 
& 
\textcolor{black}{Employs a 
{continuous diagnostics and mitigation system} through behavior profiling and analysis as 
{threat intelligence feeds} for identifying suspicious patterns, unusual
behaviors, and other 
anomalies 
}
& Near-RT RIC 
& \textcolor{black}{Network traffic and behavior patterns of user equipment (UEs)}
\\ \hline
\textcolor{black}{Secure slicing}
& 
\textcolor{black}{Dynamically adapts 
slices to which UEs are bound for protecting resources from being exploited by malicious actors; 
isolates malicious UEs, which are identified by the intrusion detection system, to 
ensure uninterrupted communication services for conforming users 
}
&  Near-RT RIC 
& \textcolor{black}{Key performance indicators related to each slice and historical data of similar networks and slice configurations}
\\ \hline

\end{tabular}%
}

\label{tab:ORANuse}
\end{table*}

\begin{figure*}[t]
    \centering
    \includegraphics[width=1.9\columnwidth, height=8.9cm]{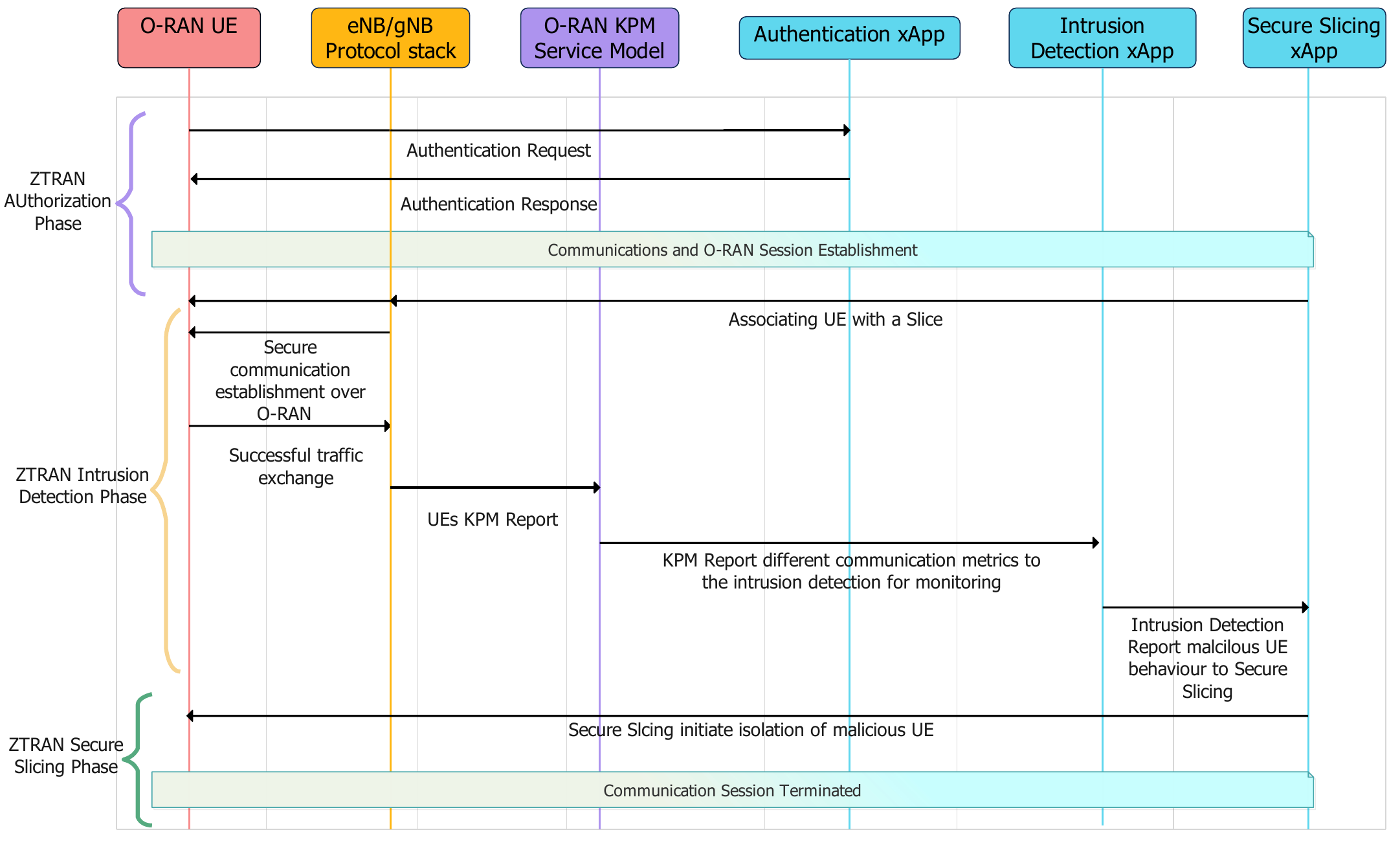}
    \vspace{-6 pt}
    \caption{\textcolor{black}{ZTRAN xApp procedures: flow of actions and O-RAN component interactions. 
    } 
    } 
    \label{fig:ZTxAppFlow}
\end{figure*}
\textcolor{black}{MFA is designed for interoperability 
across different vendor solutions. This ensures that regardless of the source of a network component, MFA can provide consistent and robust authentication, instead of relying on 
vendor-specific authentication methods.} 

\textcolor{black}{ZTRAN's authentication xApp's 
combines different authentication factors for user, RAN, and 
E2 interface identifiers. 
A random token is generated for each UE and constitutes the unique temporary UE identifier during a specific authentication transaction.
The RAN identifier is the cell identifier associated with the RAN cell to which the UE is connected. 
The E2 interface identifier is a unique identifier for the E2 connection between the RAN and the near-RT RIC. 
The UE, RAN, and E2 interface identifiers are concatenated in an encrypted format 
and provided 
to the near-RT RIC over the E2 interface. The authentication xApp 
verifies this information to ensure the authenticity of the connected UE, the reporting RAN node, and the E2 interface. 
}

\textcolor{black}{During the verification process, UEs and RAN nodes under verification are assigned only the minimum level of access and resources necessary to perform the authentication. Once the authentication xApp completes the verification process, UEs are either bound to a network slice that 
meets the resource demands or isolated to a limited bandwidth slice. This approach embodies the principle of least privilege, a core tenet of the zero trust model. Moreover, ZTRAN enforces a strict sequence of authentication: the UE cannot be authenticated until the RAN's response regarding serving this UE is authenticated first. This approach ensures that all incoming data from unauthenticated RAN nodes are ignored, minimizing the potential security risks associated with unauthorized access.} 

\textcolor{black}{As part of ZTRAN's continuous authentication strategy, slice identifiers are incorporated into the periodic authentication procedures in future authentication transactions for enhanced security. 
The slice identifier is used to verify that the resources consumed by UEs as reported by the RAN match the registered slice parameters. 
This aligns with the \textit{always verify} principle of 
zero trust security and implements the principles of the \textit{ID management} component of the zero trust architecture presented in Section~II. 
}

\subsection{ ZTRAN--Intrusion Detection xApp}
\textcolor{black}{
ZTRAN's intrusion detection system  is an xApp that 
operates continuously 
for monitoring network activities, 
examining data traffic for signs of anomalous or suspicious behavior. 
As the wireless environment and user behaviors change over time, user behavior profiling and analysis are considered for intrusion detection~\cite{IDPS_Prof}. 
ZTRAN's intrusion detection system thus continuously assesses the real time transmission behavior of the actively served UEs 
against established baselines and profiles of these UEs. 
Specifically, the intrusion detection xApp collects key performance measurements (KPMs), which include the signal-to-noise ratio, channel quality indicator, transmitted and packets, and transmission power. The mean and standard deviation of each of these KPMs are calculated and stored for different scenarios and use cases. With this information, 
the xApp creates comprehensive user behavior profiles for normal operating conditions to be used for real time intrusion detection, 
which can be based on a single KPM report or multiple successive reports provided by the RAN node over the E2 interface. 
}
  


Through continuous monitoring of user activities, {the behavior profiling and analysis unit} establishes a comprehensive understanding of normal behavior patterns within the network. It creates behavior profiles for each 
user, 
capturing typical interactions, resource accesses, and communication patterns. 
Subsequently, any unusual or suspicious activities are promptly flagged as potential threats, triggering further security measures. 
ZTRAN's intrusion detection system can be 
extended to compare the collected malicious behaviors against a database of known threat signatures or patterns, such as specific attack patterns, and enable the system to proactively recognize familiar threats.

The intrusion detection xApp is a new O-RAN microservice meant 
that provides dynamic wireless network security measures. 
It offers a customized O-RAN solution and a departure from conventional approaches, such as continuous diagnostics and mitigation, SIEM, and network and system activity logs that are commonly employed in traditional zero trust architectures. The continuous  network monitoring 
and proactive response---via secure slicing---to potential threats 
is in contrast to relying on passive monitoring and post-incident analysis. 

\subsection{ ZTRAN--Secure Slicing xApp}
\textcolor{black}{Slicing has been introduced for 5G to accommodate multiple operators, heterogeneous services, or diverse users, isolating their resources on a shared network~\cite{SS2}. Slice management and resource control can be encapsulated in an xApp~\cite{nexran}. 
We, therefore, propose extending this feature to secure slicing, considering an intruder to be a UE that has gained access to network services.}

\textcolor{black}{ZTRAN's secure slicing xApp 
defines slices that are tailored to meet specific user requirements and resource demands. It enables 
fine-grained control over 
resource allocated to UEs. 
This xApp ensures that the QoS requirements 
of authenticated users can be met within their designated {slices}, as in~\cite{9440688}. 
Once UEs are bound to slices and actively use communications services, 
continuous monitoring 
and analysis is performed by ZTRAN's intrusion detection system. 
If a UE shows anomalous behavior and is flagged by the intrusion detection algorithm, 
the secure slicing xApp bounds this UE to 
a highly restricted and isolated slice, which limits the network resources that this user has access to, preventing it from 
saturating the network. 
}
This isolation ensures that the malicious UE is denied access to the shared resource pool that it would have access to 
if secure slicing were not {employed}~\cite{ss1}. This agile approach goes beyond the static and rigid nature of conventional data access policies, enabling fine-grained resource control that can adapt and promptly address potential threats through this xApp's interactions with other xApps (intrusion detection) and the environment (RAN). 

\textcolor{black}{The secure slicing here assigns contiguous resource blocks to UEs or UE groups in a cell (O-RU), where the individual UEs are then scheduled within their respective slices~\cite{ssxapp}. The scheduler operates at the MAC layer of the O-DU and assigns resource block groups to UEs. The secure slicing process 
can distinguish between different user priorities, e.g. commercial and mission-critical users. 
Slices are not necessarily contiguous in frequency and continuous in time; they can be interlaced. 
For simplicity and without lack of generality, we assume that each UE is bound to a dedicated slice. For equal priority UEs using the same network service, such as enhanced mobile broadband, the secure slicing xApp splits the operator bandwidth in equal parts and communicates this to the RAN via the E2 interface. 
Slices can be expanded or redefined by the xApp at the granularity of radio frames (10 ms) and UEs can be unbound from one slice and bound to another, using a bitmap that is embedded in the control signal transmitted to the O-DU over the E2 interface. 
} 

\textcolor{black}{\subsection{ZTRAN--xApp Procedures and Interplay}
Table II summarizes the working principles and dependencies of ZTRAN and Fig.~\ref{fig:ZTxAppFlow} presents the workflow. 
The workflow begins with the authentication xApp. It is initiated once a new UE 
is attached to the cell. The UE sends its verification ID as part of the authentication request to the RAN which transports it to the near-RT RIC via the E2 interface. 
Within the near-RT RIC, the received authentication request is processed by ZTRAN's authentication manager that verifies the identities of the RAN, E2 interface, and UE. 
After 
successful authentication, the UE can establish communication sessions and ZTRAN's intrusion detection xApp starts monitoring its network activities. 
Periodically obtained KPM reports from the RAN are compared against the known UE profile and expected behavior for each UE. 
\textcolor{black}{If any KPM field deviates from the expected range, the presence of a potential intruder is flagged} and the secure slicing xApp is notified via the RIC message router and shared data layer. The secure slicing xApp then isolates the intruder to a dedicated network slice. 
}

\section{
 Testbed Deployment and Analysis }
\label{sec:testbed}
\begin{figure}[t]
    \centering
    \includegraphics[width=0.48\textwidth]{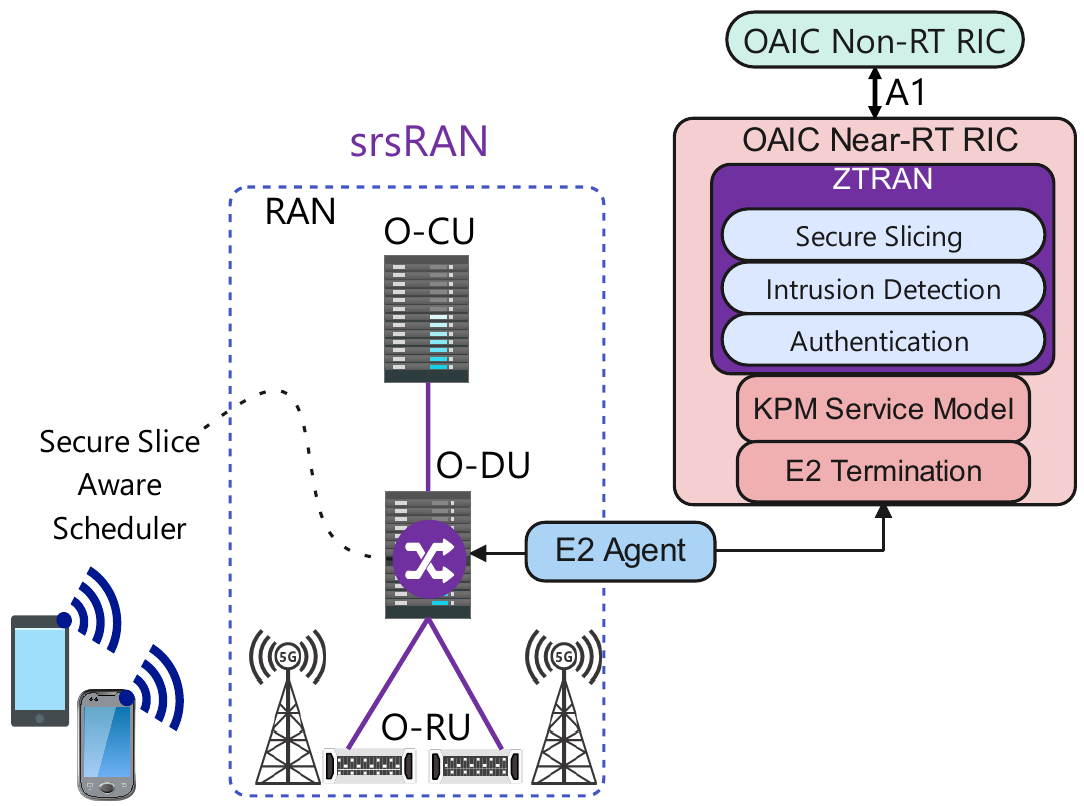}
    \caption{\textcolor{black}{The OAIC testbed implementing ZTRAN.} 
    } 
    \label{fig:OAICZTRAN}
    \vspace{-4mm}
\end{figure}
\begin{figure*}[htbp]
    \centering
    \begin{subfigure}[b]{0.99\columnwidth}
    \hspace{-125 pt}
        \includegraphics[width=2.0\textwidth, height=8.9cm]{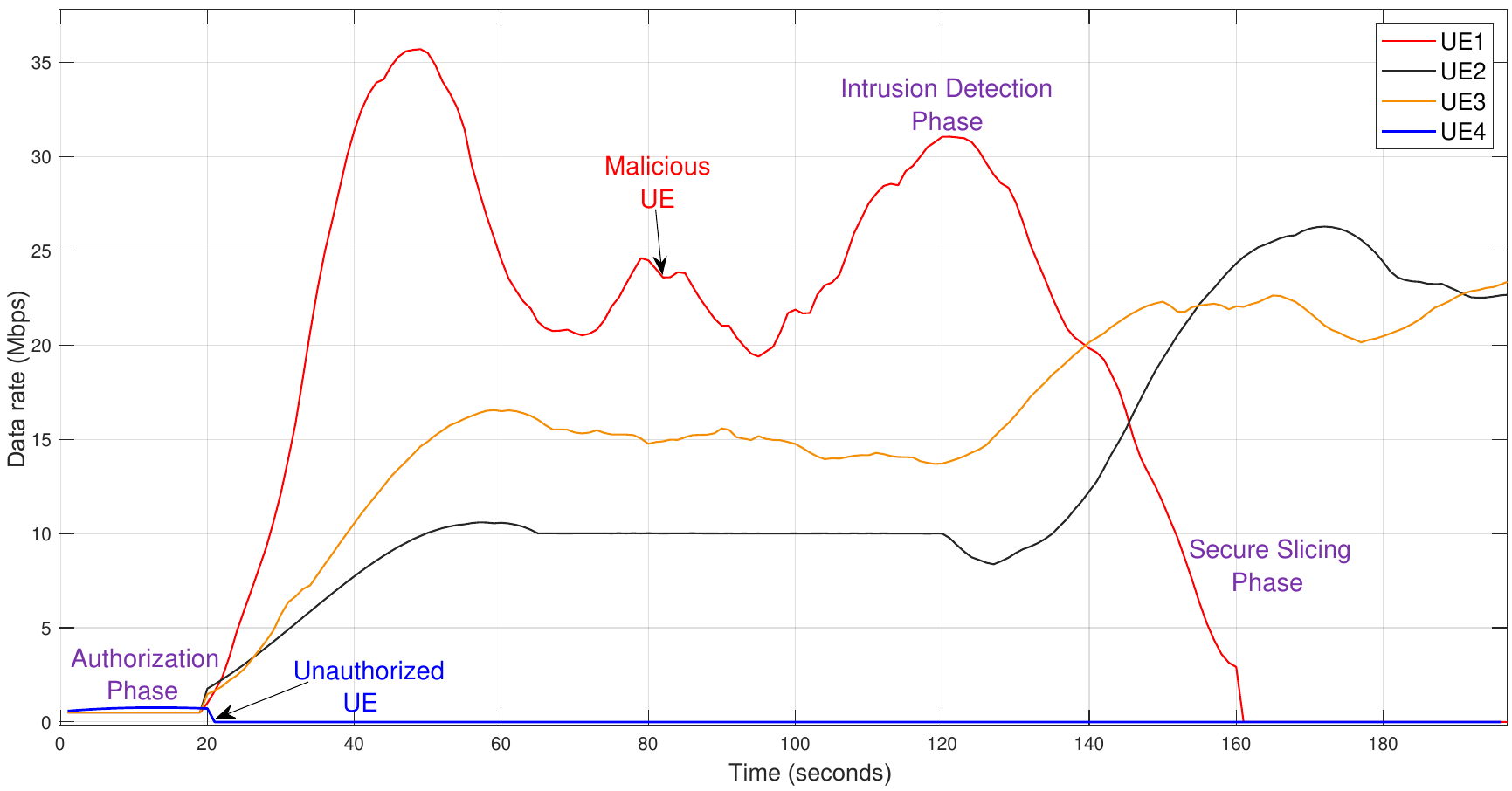}
        \caption{}
        \label{fig:ZTxApp}
    \end{subfigure}
    \hfill

    \begin{subfigure}[b]{0.5\textwidth}
        \includegraphics[width=\linewidth]{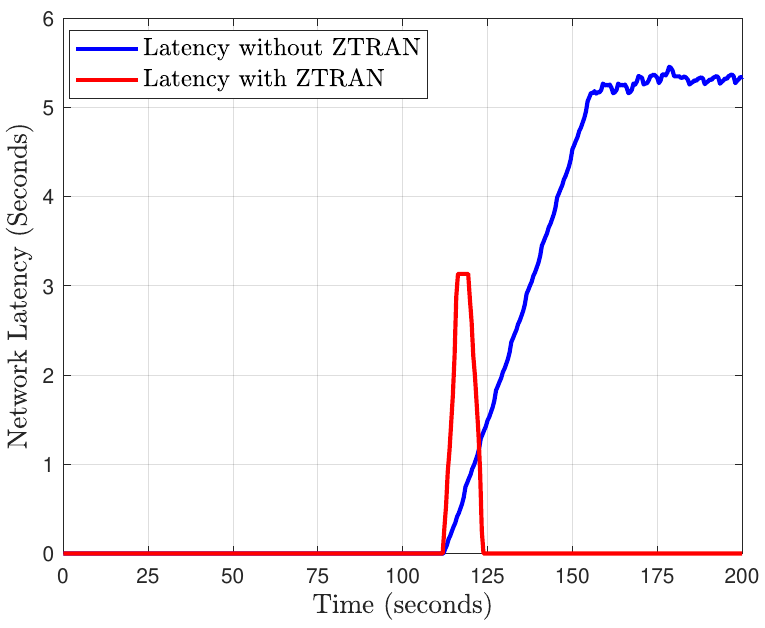}
        \caption{}
        \label{fig:latency}
    \end{subfigure}    
    \begin{subfigure}[b]{0.49\textwidth}
        \includegraphics[width=1.0\linewidth]{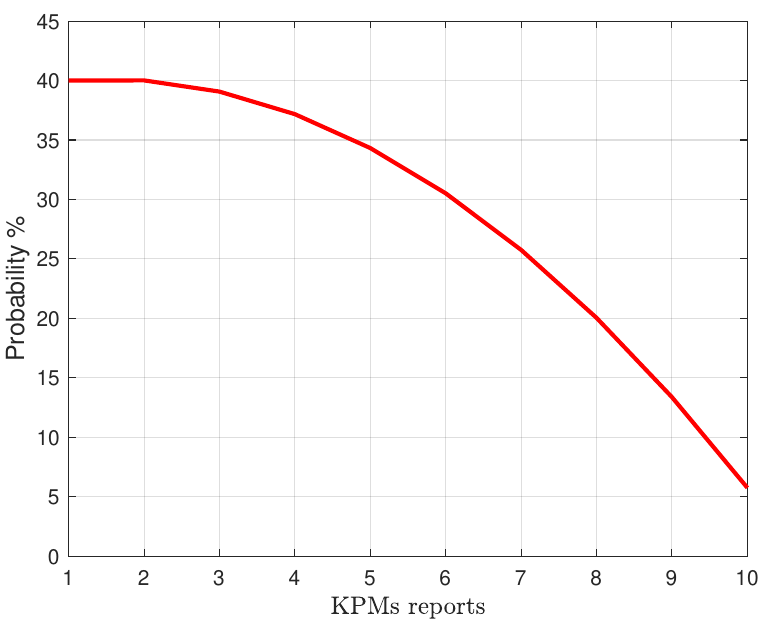}
        \caption{}
        \label{fig:FPRate}
    \end{subfigure}
    
    \caption{\textcolor{black}{Illustration of the achieved data rate of legitimate and malicious UEs served by ZTRAN (a), network latency performance without and with ZTRAN (b), and false positive rate performance of the intrusion detection xApp (c).} 
    }
    \label{results}
\end{figure*}
\textcolor{black}{
\textcolor{black}{Implementation of ZTRAN and experimental verification and analysis is enabled by the OAIC platform. OAIC 
is an open-source O-RAN research platform that enables rapid prototyping and validation on software radio testbeds~\cite{Aly_ORAN2}.} \textcolor{black}{It employs srsRAN's 4G/5G software radio software for the UEs and RAN and implements O-RAN Alliance's near-RT RIC, E2 interface protocols and end points, and service models.} 
\textcolor{black}{Figure~\ref{fig:OAICZTRAN} shows the components involved in the experiments of this paper, where zero message queue is employed.} 
}



\textcolor{black}{OAIC implements the near-RT RIC interfaces and processes provided by the O-RAN Software Community to deploy and execute xApps and enable interactions with 
the RAN. The near-RT RIC is deployed on a workstation with a 10-core Intel Xeon W-2255 processor with 128 GB of RAM and 1 TB of storage capacity. Virtualization is implemented at the operating system level using Kubernetes as the virtual infrastructure manager. Each layer of the O-RAN stack implements 
through a microservice-based architecture, where an application is decomposed into several parts running in their own lightweight environment. This makes the system more robust and easier to maintain, scale, test, and employ for new use cases. }

\textcolor{black}{Figure~\ref{fig:ZTxApp} presents the data rate performance of four UEs that aim to access O-RAN resources. 
\textcolor{black}{
All UEs 
are initially attached to a 20 MHz cell sharing 100 physical resource blocks
(PRBs). 
In this configuration, any UE can request up to 100 PRBs. However, the resources accessible by a UE 
and the achievable throughput may change 
over time and are a function of the security measures taken by ZTRAN. 
} 
First the UEs go through the MFA scheme where each user is required to present a combination of IDs as previously discussed. 
This authentication process verifies the identity of each UE and ensures that only authorized and authenticated UEs, as verified by the authentication xApp, can gain access to O-RAN resources.} 

\textcolor{black}{According to Fig.~\ref{fig:ZTxApp}, UE1-UE3 have presented the right credentials to gain access to O-RAN resources while UE4 has been denied access.} Once the 
authentication phase is completed, the authorized UEs can access O-RAN resources, and initiate active communication 
sessions. 
While the data transmissions are active, ZTRAN's 
intrusion detection system 
monitors 
user behaviors based on several metrics provided by the 
KPM service model. This includes evaluating parameters such as packet count and network traffic patterns. 
\textcolor{black}{
The intrusion detection xApp considers the normal data rate to be between 10 and 20 Mbps. There is one UE that exceeds this range as shown in Fig.~4a. 
} \textcolor{black}{By processing the collected KPMs, 
ZTRAN's intrusion detection system detects massive uplink transmission packets generated by UE1. It concludes that a potential flooding attack is taking place that could overwhelm the network and 
affect the network service provisioning to other UEs.} By detecting this threat, ZTRAN activates secure slicing, 
which promptly isolates the malicious UE to a dedicated slice 
followed by 
throttling its data rate. 
Consequently, the resources initially consumed by the malicious UE are reallocated 
to the other two UEs, as illustrated in Fig.~\ref{fig:ZTxApp}.

\textcolor{black}{We evaluate the case of a default network configuration without and with 
ZTRAN deployment 
for three UEs, two of them legitimate and one malicious. Figure~4b illustrates the network latency results. 
In the default network configuration without ZTRAN 
the consequences are a notable degradation in network latency performance. Normal network latency on average is around $10$~ms which increases to over $5$~s when a malicious UE is present and is flooding the network with requests. 
When employing ZTRAN, the latency increase 
is temporary and the secure slicing effectively removes the malicious user from affecting the network. 
ZTRAN limits the network latency increase 
to 6.6\% of the time it was evaluated as opposed to 
50\% of the time without ZTRAN.}

\textcolor{black}{
Figure~\ref{fig:FPRate} shows the false positive probability 
over the number of collected KPM reports. 
These reports are delivered by the RAN over the E2 interface. ZTRAN's intrusion detection xApp 
uses them for user profiling and analysis. 
The false positive metric measures the rate at which benign activities are incorrectly classified as malicious. We observe that as more KPM reports are processed, the xApp is able to more accurately distinguish between normal and malicious activities. 
} 

\textcolor{black}{
The complexity of ZTRAN processes is of the order of $O(n + k + 1)$, where $n$ corresponds to the factors of the MFA mechanism and $k$ to the number of KPMs being processed for intrusion detection. Secure slicing updates access controls, adjusts resource allocations, and isolates intruders to a dedicated slice; {it is thus independent on the number of intruders}
.}

\section{{\textcolor{black}{Open Issues and R\&D Directions}} 
}
\label{sec:capabilities}
\subsection{\textcolor{black}{AI-Enabled ZTRAN}} 
Incorporating AI technologies offers a compelling opportunity to enhance the functionalities of ZTRAN. 
While the existing methods demonstrate accurate classifications, their reliance on static metrics limits their adaptability in dynamic deployment scenarios. By harnessing AI algorithms, these subsystems can continuously learn and evolve, adapting swiftly to changing network conditions and emerging attack patterns. The AI-powered components actively gather and analyze real-time network data, refining their decision-making capabilities and enabling them to detect and mitigate security threats effectively. 
\textcolor{black}{While O-RAN provides native support for AI controllers, further research is needed to evaluate the use of AI for O-RAN security services and the security vulnerabilities of AI models and processes.} 

\subsection{\textcolor{black}{Onboarding of 
xApps and Database Access Control}} \textcolor{black}{Onboarding untrusted third-party xApps 
in the near-RT RIC carries similar security risks as a malicious or poorly configured xApp. Untrusted xApps may weak API protection, excessive service exposure, and may capture sensitive user information, among other security vulnerabilities. Moreover, xApps currently have full access to the near-RT RIC database regardless of their actual need. Employing OAuth 2.0 and Role-Based Access Control (RBAC) can ensure authorized access to the database through tokens that are assigned based on roles. Proper verification and access control mechanisms are essential to mitigate these threats. Potential solutions include verifying digital signatures from trusted service providers and solution providers, confirming membership in a trusted list of providers, and checking certificate revocation status. 
We recommend building on zero trust security principles to research and develop rigorous authentication and validation checks that ensure that only authorized and unaltered xApps are deployed on the near-RT RIC platform and that xApps get only the necessary access rights without compromising performance.}

\vspace{-20 pt}
\textcolor{black}{\subsection{xApp Conflict Resolution}  
The near-RT RIC concurrently executes several xApps, each providing a specific microservice. These xApps may have been developed independently by different groups and tested in isolation. Conflicts may arise if more than one microservice 
tries to control the same resource or resources that depend on each other. 
Conflicting decisions on resource allocation among xApps may lead to suboptimal performance, inefficiencies, or even degradation of service quality. Resolving these conflicts while ensuring that each xApp meets its optimization goals poses a significant open challenge. 
Coordination mechanisms may include dynamic resource negotiation, intelligent prioritization schemes, and conflict resolution strategies to ensure that the concurrent execution of xApps does not compromise the efficiency and effectiveness of the network in terms of stability, latency, and throughput, among other metrics. 
} 

\vspace{-20 pt}
\textcolor{black}{\subsection{
Contextual Awareness} 
ZTRAN's intrusion detection xApp 
performs behavior profiling and analysis. Profiling focuses on the behavioral aspects of users and devices, but it may lack contextual awareness of the broader network environment. Intruders may exploit vulnerabilities that are context-dependent and not reflected in individual behavior profiles, necessitating additional contextual information for accurate detection. Depending solely on profiling has led to the need to accumulate ten KPM reports for ensuring high detection and low false positive rates. Therefore, additional threat intelligence fees should be considered 
to enrich the contextual information available to the intrusion detection xApp. Furthermore, we recommend exploring 
ML and predictive analytics to anticipate contextual changes and potential threats and develop user-centric security policies that consider both individual behaviors and contextual factors. }


\section{Conclusions}
\label{sec:conclusions}
\textcolor{black}{This paper argues for the application of zero trust principles for improving O-RAN security. 
We introduce ZTRAN, which offers O-RAN security tools enabling R\&D on open, virtualized, intelligent, and secure advanced wireless networks. 
ZTRAN implements new microservices that are deployed on the 
near-RT RIC.  
It currently offers user service authentication, intrusion detection, and secure slicing to verify, monitor, and control end users based on the KPM reports it receives from the RAN over the E2 interface. 
Experimental results reveal the improved throughput and latency performance of legitimate users by timely identifying and isolating intruders. This demonstrates how ZTRAN's intrusion detection and secure slicing microservices operate effectively and in concert. 
The implementation on the OAIC platform shows ZTRAN's compatibility with O-RAN Alliance's network architecture and O-RAN Software Community's containerized near-RT RIC. 
The experimental results and findings of this paper provide new opportunities for R\&D, encouraging the design, development, and testing of xApps for offering diverse O-RAN security services. 
}

\section*{Acknowledgement}
This work was supported in part by the National Science Foundation (NSF) under grant number 2120442 and by NSF and Office of the Under Secretary of Defense (OUSD) – Research and Engineering, under Grant ITE2326898, as part of the NSF Convergence Accelerator Track G: Securely Operating Through 5G Infrastructure Program.

\balance

\bibliographystyle{IEEEtran}
\bibliography{main}

\vspace{-5 pt}
\section*{Biographies}
\footnotesize
\vspace{0.2cm}
\noindent
\textbf{Aly Sabri Abdalla} (asa298@msstate.edu)
is an Assistant Research Professor in the Department of Electrical and Computer Engineering at Mississippi State University, Starkville, MS, USA. His research interests are on wireless communication and networking, software radio, spectrum sharing, wireless testbeds and testing, and wireless security with application to mission-critical communications, open radio access network (O-RAN), unmanned aerial vehicles (UAVs), and reconfigurable intelligent surfaces (RISs).

\vspace{0.2cm}
\noindent
\textbf{Joshua Moore} (jjm702@msstate.edu)
is a PhD student in the Department of Electrical and Computer Engineering at Mississippi State University, Starkville, MS, USA. His research interests include O-RAN, 5G/next-G communications, and RAN Management and Orchestration.

\vspace{0.2cm}
\noindent
\textbf{Nisha Adhikari} (na731@msstate.edu)
is a Undergraduate student in the Department of Electrical and Computer Engineering at Mississippi State University, Starkville, MS, USA. Her research interests include AI and software development in ORAN for wireless communications.

\vspace{0.2cm}
\noindent

\vspace{0.2cm}
\noindent
\textbf{Vuk Marojevic} (vuk.marojevic@msstate.edu) is an associate professor in electrical and computer engineering at Mississippi State University, Starkville, MS, USA. His research interests include resource management, vehicle-to-everything communications and wireless security with application to cellular communications, mission-critical networks, and unmanned aircraft systems.

\end{document}